\documentclass[epj]{webofc}
\usepackage[utf8]{inputenc}
\usepackage[varg]{txfonts}   
\usepackage{booktabs}
\usepackage{xcolor}
\definecolor{darkred}{rgb}{0.4,0.0,0.0}
\definecolor{darkgreen}{rgb}{0.0,0.4,0.0}
\definecolor{darkblue}{rgb}{0.0,0.0,0.4}
\usepackage[bookmarks,linktocpage,colorlinks,
    linkcolor = darkred,
    urlcolor  = darkblue,
    citecolor = darkgreen]{hyperref}

\usepackage{subfigure}
\wocname{EPJ Web of Conferences}
\woctitle{Lattice2017}
\begin{document}
\begin{flushright}
KEK-CP-363, RBRC-1256
\end{flushright}
%
\selectlanguage{english}
\title{%
Axial $U(1)$ symmetry at high temperature in 2-flavor lattice QCD
}
\author{%
\firstname{Kei} \lastname{Suzuki}\inst{1}\fnsep\thanks{Speaker, \email{kei.suzuki@kek.jp}} \and
\firstname{Sinya} \lastname{Aoki}\inst{2} \and
\firstname{Yasumichi} \lastname{Aoki}\inst{1,3} \and
\firstname{Guido}  \lastname{Cossu}\inst{4} \and
\firstname{Hidenori}  \lastname{Fukaya}\inst{5} \and
\firstname{Shoji}  \lastname{Hashimoto}\inst{1,6} \ (JLQCD Collaboration)
}
\institute{%
KEK Theory Center, High Energy Accelerator Research Organization (KEK), Tsukuba 305-0801, Japan
\and
Center for Gravitational Physics, Yukawa Institute for Theoretical Physics, Kyoto 606-8502, Japan
\and
RIKEN BNL Research Center, Brookhaven National Laboratory, Upton NY 11973, USA
\and
School of Physics and Astronomy, The University of Edinburgh, Edinburgh EH9 3JZ, United Kingdom
\and
Department of Physics, Osaka University, Toyonaka 560-0043, Japan
\and
School of High Energy Accelerator Science, The Graduate University for Advanced Studies (Sokendai), Tsukuba 305-0801, Japan
}
\abstract{%
We investigate the axial $U(1)_A$ symmetry breaking above the critical temperature in two-flavor lattice QCD.
The ensembles are generated with dynamical M\"obius domain-wall or reweighted overlap fermions.
The $U(1)_A$ susceptibility is extracted from the low-modes spectrum of the overlap Dirac eigenvalues.
We show the quark mass and temperature dependences of $U(1)_A$ susceptibility.
Our results at $T=220 \, \mathrm{MeV}$ imply that the $U(1)_A$ symmetry is restored in the chiral limit.
Its coincidence with vanishing topological susceptibility is observed.
}
\maketitle
\section{Introduction}\label{sec-1}
In quantum chromodynamics (QCD) at low temperature, the axial $U(1)_A$ symmetry is violated by the quantum (chiral) anomaly, which is the origin of much heavier $\eta^\prime$ meson than other pseudoscalar mesons.
The $U(1)_A$ symmetry breaking is closely related to topological excitations of the background gauge field, such as instantons.
As an observables to characterize the $U(1)_A$ symmetry breaking, the $U(1)_A$ susceptibility $\Delta_{\pi - \delta}$ is defined by the difference between the correlators of isovector-pseudoscalar ($\pi^a \equiv i \bar{\psi} \tau^a \gamma_5 \psi$) and isovector-scalar ($\delta^a \equiv \bar{\psi} \tau^a \psi$) operators: 
\begin{equation}
\Delta_{\pi-\delta} \equiv \chi_\pi - \chi_\delta \equiv \int d^4x \langle \pi^a(x) \pi^a(0) - \delta^a (x) \delta^a(0) \rangle, \label{eq:Delta_def}
\end{equation}
where $a$ is the isospin index. (We consider the theory with two degenerate quark flavors.)

Above the critical temperature, $T>T_c$, while the (spontaneously broken) chiral symmetry is known to be restored, the restoration/violation of the $U(1)_A$ symmetry is a long standing problem, which has been studied using in analytic methods \cite{Cohen:1996ng,Aoki:2012yj,Kanazawa:2015xna} and effective theories \cite{Gross:1980br,Pisarski:1983ms} as well as lattice QCD simulations at $N_f=2$ \cite{Cossu:2013uua,Chiu:2013wwa,Brandt:2016daq,Tomiya:2016jwr} and $N_f=2+1$ \cite{Bazavov:2012qja,Buchoff:2013nra,Bhattacharya:2014ara,Dick:2015twa}.
In Ref.~\cite{Cohen:1996ng}, Cohen claimed that the $U(1)_A$ symmetry of massless $N_f=2$ QCD is restored when the contributions from the zero modes of Dirac eigenvalues (i.e. nontrivial topological sectors) can be ignored.
As a result, the mesonic correlators for $\pi$, $\sigma$, $\delta$, and $\eta$ channels become degenerate.

\begin{figure}[tb!]
    \begin{minipage}[t]{1.0\columnwidth}
        \begin{center}
            \includegraphics[clip, width=1.0\columnwidth]{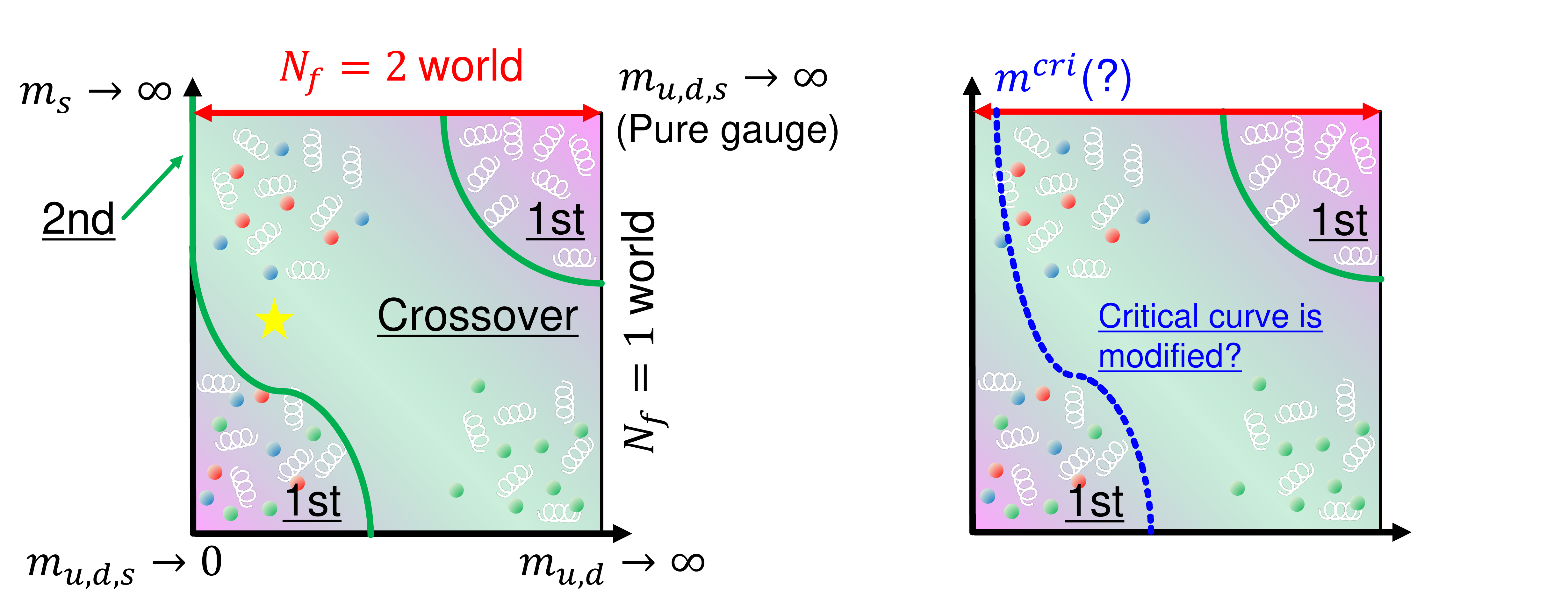}
        \end{center}
    \end{minipage}
    \caption{Phase diagrams of QCD varying up and down quark mass $m_{u,d}$ and strange quark mass $m_s$.
Left: The conventional diagram.
Right: A possible diagram when the $U(1)_A$ is restored above $T_c$, which is suggested in Ref.~\cite{Aoki:2012yj}.}
    \label{Fig-col}
\end{figure}

More recently, the authors of Ref.~\cite{Aoki:2012yj} proved that the violation of the $U(1)_A$ symmetry of massless $N_f=2$ QCD becomes invisible under some plausible assumptions including the analyticity of the Dirac spectral density near the origin (for an alternative proof, see Ref.~\cite{Kanazawa:2015xna}).
They even suggested a possible modification of the phase diagram in the space of up or down quark mass $m_{u,d}$ and strange quark mass $m_s$ (the so-called Columbia plot, see Fig.~\ref{Fig-col}), according to the argument based on the effective theory \cite{Pisarski:1983ms}.
Because of the restoration of $U(1)_A$ symmetry in the chiral limit for $N_f=2$, the chiral phase transition at $m_{u,d}=0$ may be first-order rather than the (usually expected) second-order~\footnote{Strictly speaking, second-order belonging to universality classes other than the (usually expected) $O(4)$ class is also possible.}.
If this is the case, a nonzero ``critical mass,'' $m_{u,d}^\mathrm{cri}$, appears, which separates the first-order region $m<m_{u,d}^\mathrm{cri}$ and the crossover region $m>m_{u,d}^\mathrm{cri}$ for $N_f=2$.
The existence of such a critical mass can also affect the phase structure of $N_f=3$ QCD.

In the previous works by the JLQCD Collaboration \cite{Cossu:2013uua,Tomiya:2016jwr}, we observed restored $U(1)_A$ symmetry above $T_c$ in $N_f=2$ lattice QCD.
Because the $U(1)_A$ susceptibility is sensitive to the chiral symmetry on the lattice, we used the lattice fermions that maintain the chiral symmetry, i.e. the overlap (OV) or domain-wall (DW) fermion formalism.
In Ref.~\cite{Cossu:2013uua}, the $U(1)_A$ symmetry was investigated from the Dirac spectrum on gauge configurations generated with the dynamical overlap fermions under a fixed global topology, $Q=0$.
In Ref.~\cite{Tomiya:2016jwr}, the gauge configurations are generated with dynamical M\"obius domain-wall fermions \cite{Brower:2005qw,Brower:2012vk}.
Since the Ginsparg-Wilson (GW) relation \cite{Ginsparg:1981bj} for the M\"obius domain-wall fermion is slightly violated especially for larger lattice spacings \cite{Cossu:2015kfa}, we applied the domain-wall/overlap reweighting \cite{Tomiya:2016jwr}, where an observable on the gauge ensembles generated with dynamical domain-wall fermions is reweighted to that of overlap fermions.
The exact chiral symmetry is thus realized without sacrificing the topology sampling in Ref.~\cite{Tomiya:2016jwr}.

In these proceedings, we further study the $U(1)_A$ symmetry above $T_c$ in $N_f=2$ lattice QCD.
Our numerical setup is updated, compared to that in the previous paper \cite{Tomiya:2016jwr}.
A finer lattice spacing, $1/a=2.64 \, \mathrm{GeV}$ ($a \sim 0.075 \, \mathrm{fm}$), is used, which improves the GW relation of the M\"obius domain-wall fermion.

\section{Simulation setup}\label{sec-2}

\subsection{$U(1)_A$ susceptibility on the lattice}\label{subsec-2-1}
The $U(1)_A$ susceptibility (\ref{eq:Delta_def}) can be written in terms of the spectral density $\rho(\lambda)$ of Dirac eigenvalues $\lambda$ for fermions with a mass $m$.
In the continuum theory, it reads
\begin{equation}
\Delta_{\pi-\delta} = \int_0^\infty d\lambda\,\rho(\lambda) \frac{2m^2}{(\lambda^2+m^2)^2}, \label{eq:Delta_cont}
\end{equation}
where the eigenvalue spectrum is defined by $\rho(\lambda)=(1/V)\langle\sum_{\lambda'}\delta(\lambda-\lambda')\rangle$ and the four-dimensional volume is $V=L^3\times L_t$.

On the lattice with overlap fermion formulation, the $U(1)_A$ susceptibility is given by \cite{Cossu:2015kfa}
\begin{equation}
\Delta_{\pi-\delta}^{\mathrm{ov}} =  \frac{1}{V(1-m^2)^2} \left< \sum_i \frac{2m^2(1-\lambda_i^{(\mathrm{ov},m)2})^2}{\lambda_i^{(\mathrm{ov},m)4}} \right> , \label{eq:Delta_ov}
\end{equation}
where $\lambda_i^{(\mathrm{ov},m)}$ is the $i$-th eigenvalue of the massive overlap Dirac operator, and we set the lattice spacing $a=1$.
When the GW relation is violated, there are additional terms in Eq.~(\ref{eq:Delta_ov}) \cite{Cossu:2015kfa}.
Eq.~(\ref{eq:Delta_ov}) includes the effect of nontrivial topological sectors as chiral zero modes: $\lambda_i^{(\mathrm{ov},m)} \approx \pm m$, where ``$\approx$'' implies possible small violation of the GW relation in our simulations (If the GW relation is exact, then $\lambda_i^{(\mathrm{ov},m)} = \pm m$).
After such zero modes are subtracted from Eq.~(\ref{eq:Delta_ov}), we define an improved estimate of the $U(1)_A$ susceptibility:
\begin{equation}
\bar{\Delta}_{\pi-\delta}^{\mathrm{ov}} \equiv \Delta_{\pi-\delta}^{\mathrm{ov}} - \frac{1}{V(1-m^2)^2} \left< \sum_{0-mode} \frac{2m^2(1-\lambda_i^{(\mathrm{ov},m)2})^2}{\lambda_i^{(\mathrm{ov},m)4}} \right> . \label{eq:Delta_bar}
\end{equation}
The subtraction of chiral zero modes in Eq.~(\ref{eq:Delta_bar}) is justified as follows \cite{Aoki:2012yj}.
If the GW relation is exact, the second term of Eq.~(\ref{eq:Delta_bar}) can be written as $2N_0/Vm^2$, where $N_0$ is the number of chiral zero modes.
$\langle N_0^2 \rangle$ is expected to scale as $O(V)$, so that $N_0/V$ as $O(1/\sqrt{V})$.
Therefore, the contribution from exact zero modes should vanish in the thermodynamic limit: $N_0/V \to 0$ as $V \to \infty$.

The overlap-Dirac eigenvalues $\lambda_i^{(\mathrm{ov},m)}$ measured on the M\"obius domain-wall fermion ensembles may include fictitious (zero and nearzero) modes induced by the partially quenched approximation \cite{Tomiya:2016jwr}.
They come from a mismatch between the eigenvalues and the fermion determinant, which determines the Boltzmann factor for the particular gauge configuration.
After the DW/OV reweighting, the mismatch is resolved by giving negligible small reweighting factor for the gauge configuration that suffer from the fictitious modes.
In Section \ref{sec-3}, we compare $\lambda_i^{(\mathrm{ov},m)}$ and $\bar{\Delta}_{\pi-\delta}^{\mathrm{ov}}$ on the DW ensembles with those on the OV ensembles.

\subsection{Numerical setup}\label{subsec-2-2}
The parameters in our numerical study are summarized in Table \ref{Tab:param}.
We use the lattice with the spatial size $L=32$ and the temporal length $L_t =12$ and $8$ corresponding to $T=220$ and $330\, \mathrm{MeV}$, respectively, at the lattice spacing, $1/a=2.64 \, \mathrm{GeV}$ ($a \sim 0.075 \, \mathrm{fm}$).
We take quark masses, $am=0.001-0.01$ ($2.64-26.4 \, \mathrm{MeV}$).

\begin{table}[t!]
  \small
  \centering
\caption{Numerical parameters in lattice simulations.
$L^3 \times L_t $, $L_s$, $\beta$, $a$, and $m$ are the lattice size, length of the fifth dimension in the M\"obius domain-wall fermion,
gauge coupling, lattice spacing, and quark mass, respectively.
}
\begin{tabular}{cccccc}
\hline\hline
$L^3 \times L_t $ & $L_s$ & $\beta$ & $a$ [fm] & $T$ [MeV] & $am$ \\
\hline
$32^3 \times 12$ & 16 & 4.30 & 0.075 & 220 & 0.001   \\
$32^3 \times 12$ & 16 & 4.30 & 0.075 & 220 & 0.0025  \\
$32^3 \times 12$ & 16 & 4.30 & 0.075 & 220 & 0.00375 \\
$32^3 \times 12$ & 16 & 4.30 & 0.075 & 220 & 0.005   \\
$32^3 \times 12$ & 16 & 4.30 & 0.075 & 220 & 0.01    \\
\hline
$32^3 \times 8$  & 12 & 4.30 & 0.075 & 330 & 0.001   \\
$32^3 \times 8$  & 12 & 4.30 & 0.075 & 330 & 0.005   \\
$32^3 \times 8$  & 12 & 4.30 & 0.075 & 330 & 0.01    \\
\hline\hline
\end{tabular}
\label{Tab:param}
\end{table}

We use the tree-level Symanzik improved gauge action.
For the fermion part, the smeared M\"obius domain-wall fermion is applied.
The four dimensional effective operator of the M\"obius DW fermion with a mass $m$ is given as \cite{Brower:2005qw,Brower:2012vk}
\begin{equation}
D_{DW}^{4D}(m) = \frac{1+m}{2} + \frac{1-m}{2} \gamma_5 \mathrm{sgn} (H_M),
\end{equation}
where ``sgn'' is the matrix sign function.
It is approximated as $\tanh (L_s \tanh^{-1}(H_M))$ with the length of the fifth dimension, $L_s$.
It becomes the exact sign function in the limit $L_s \to \infty$.
The kernel operator $H_M$ is \cite{Brower:2005qw,Brower:2012vk}
\begin{equation}
H_M=\gamma_5 \frac{\alpha D_W}{2+D_W}, \label{eq:kernel}
\end{equation}
where $D_W$ is the Wilson-Dirac operator with a large negative mass $-1/a$, and we set the scale parameter $\alpha$ to $2$.

In this study, we focus on the low-lying overlap Dirac eigenmodes to evaluate the $U(1)_A$ susceptibility from Eq.~(\ref{eq:Delta_ov}) or (\ref{eq:Delta_bar}), but the polar approximation for the sign function with $L_s$ used in our simulations is insufficient for the low modes to satisfy the GW relation \cite{Cossu:2015kfa}.
Instead, we exactly calculate the sign function for the lowest eigenmodes of $H_M$ below some threshold $\lambda_{th}^M$.
Namely, we construct the overlap Dirac operator as follows \cite{Fukaya:2013vka,Tomiya:2016jwr}:
\begin{equation}
D_{\mathrm{ov}}(m) = \sum_{ |\lambda_i^M|<\lambda_{th}^M} \left[ \frac{1+m}{2} + \frac{1-m}{2} \gamma_5 \mathrm{sgn} (\lambda_i^M) \right] | \lambda_i^M \rangle \langle \lambda_i^M | + D_{DW}^{4D}(m) \left[ 1 - \sum_{\lambda_{th}^M < |\lambda_i^M| } | \lambda_i^M \rangle \langle \lambda_i^M | \right],
\end{equation}
where $\lambda_i^M$ is the $i$-th eigenvalue of the kernel operator (\ref{eq:kernel}).
In this form, the first and second terms correspond to the (separately treated) lower and higher modes, respectively.
Since the approximation of the sign function is sufficiently precise for large $\lambda_i^M$'s, the operator constructed in this way satisfies the GW relation nearly exactly.

An observable $\mathcal{O}$ measured on the M\"obius DW fermion ensembles may be affected by the violation of the GW relation, but they can be transformed to those on the OV fermion ensembles by the DW/OV reweighting \cite{Tomiya:2016jwr}:
\begin{equation}
\langle \mathcal{O} \rangle_{\mathrm{ov}} = \frac{ \langle \mathcal{O} R\rangle_{\mathrm{DW}} }{ \langle R\rangle_{\mathrm{DW}}},
\end{equation}
where $\langle \cdots \rangle_{\mathrm{DW}}$ and $\langle \cdots \rangle_{\mathrm{ov}}$ are the ensemble average with the M\"obius DW and reweighted OV fermions, respectively.
$R$ is the reweighting factor stochastically estimated on the M\"obius DW ensembles.

\section{Preliminary results}\label{sec-3}

\begin{figure}[t!]
    \centering
    \begin{minipage}[t]{1.0\columnwidth}  
            \includegraphics[clip, width=1.0\columnwidth]{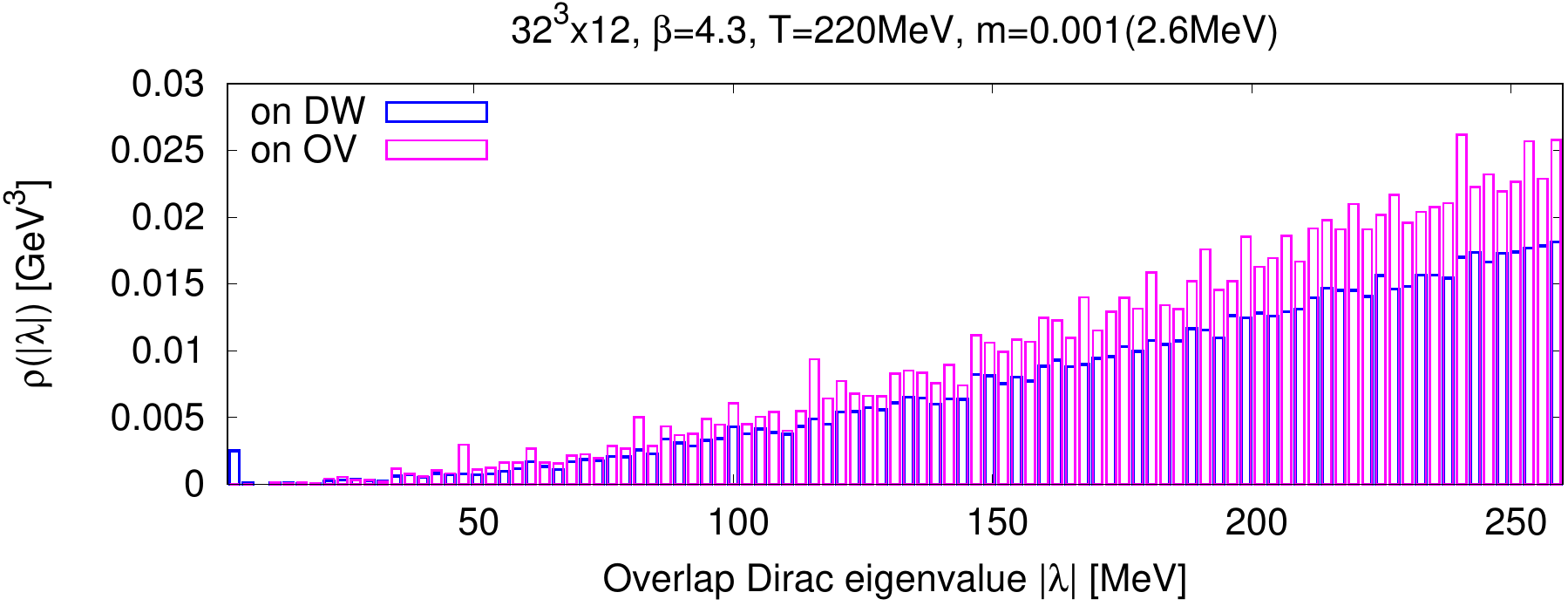}       
    \end{minipage}
    \begin{minipage}[t]{1.0\columnwidth}
            \includegraphics[clip, width=1.0\columnwidth]{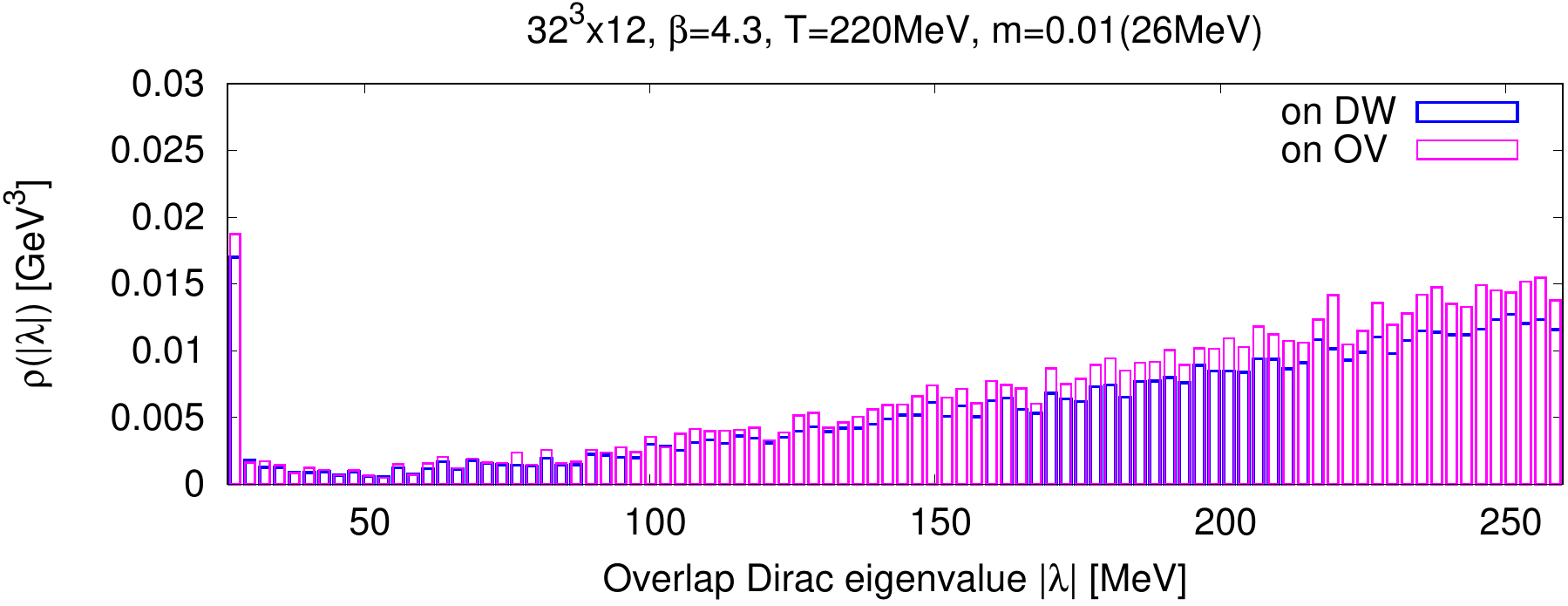}
    \end{minipage}
    \caption{Spectral density $\rho(|\lambda|)$ for overlap Dirac eigenvalues $\lambda$ at $T=220 \, \mathrm{MeV}$.
Upper panel: $m=2.6 \, \mathrm{MeV}$.
Lower panel: $m=26 \, \mathrm{MeV}$.
Blue and magenta bins correspond to the spectra on the original M\"obius domain-wall (DW) and reweighted overlap (OV) fermion ensembles, respectively.
 }
    \label{fig-1}
\end{figure}

\subsection{Spectral density of overlap Dirac eigenvalues}\label{subsec-3-1}
In Fig.~\ref{fig-1}, we show the spectral density $\rho(|\lambda|)$ of the overlap Dirac eigenvalues $\lambda$ calculated on both the M\"obius domain-wall and reweighted overlap fermion ensembles at $T=220 \, \mathrm{MeV}$, which is above $T_c$.
For the small quark mass (the upper panel in Fig.~\ref{fig-1}), we find that the low modes are suppressed (or a ``gap'' opens) due to the thermal effect.
Then, one can clearly distinguish the zero modes from other higher modes.
From Eq.~(\ref{eq:Delta_bar}), since the finite value of $\bar{\Delta}_{\pi-\delta}^{\mathrm{ov}}$ comes from the nonzero low modes, such suppression of the low modes is expected to decrease the value of $\bar{\Delta}_{\pi-\delta}^{\mathrm{ov}}$.
In addition, we note that the zero modes in the spectrum calculated on the DW fermion ensemble (shown by blue bins) are artifacts caused by the disagreement between the valence and sea quarks (partially quenched approximation).
After the reweighting (shown by magenta bins), such artificial zero modes disappear (see the leftmost bin).

For the large quark mass (the lower panel in Fig.~\ref{fig-1}), the gap on the spectra is closed, and we cannot clearly separate the zero modes from finite modes.
We expect that, the contributions from the nonzero modes lead to a finite value of $\bar{\Delta}_{\pi-\delta}^{\mathrm{ov}}$.
Furthermore, in this case, the zero modes on the DW ensemble survive after the DW/OV reweighting, which means that such zero modes are physical, and they can be related to nonzero topological susceptibility.

\begin{figure}[t!]
    \begin{minipage}[t]{0.5\columnwidth}
        \begin{center}
            \includegraphics[clip, width=1.0\columnwidth]{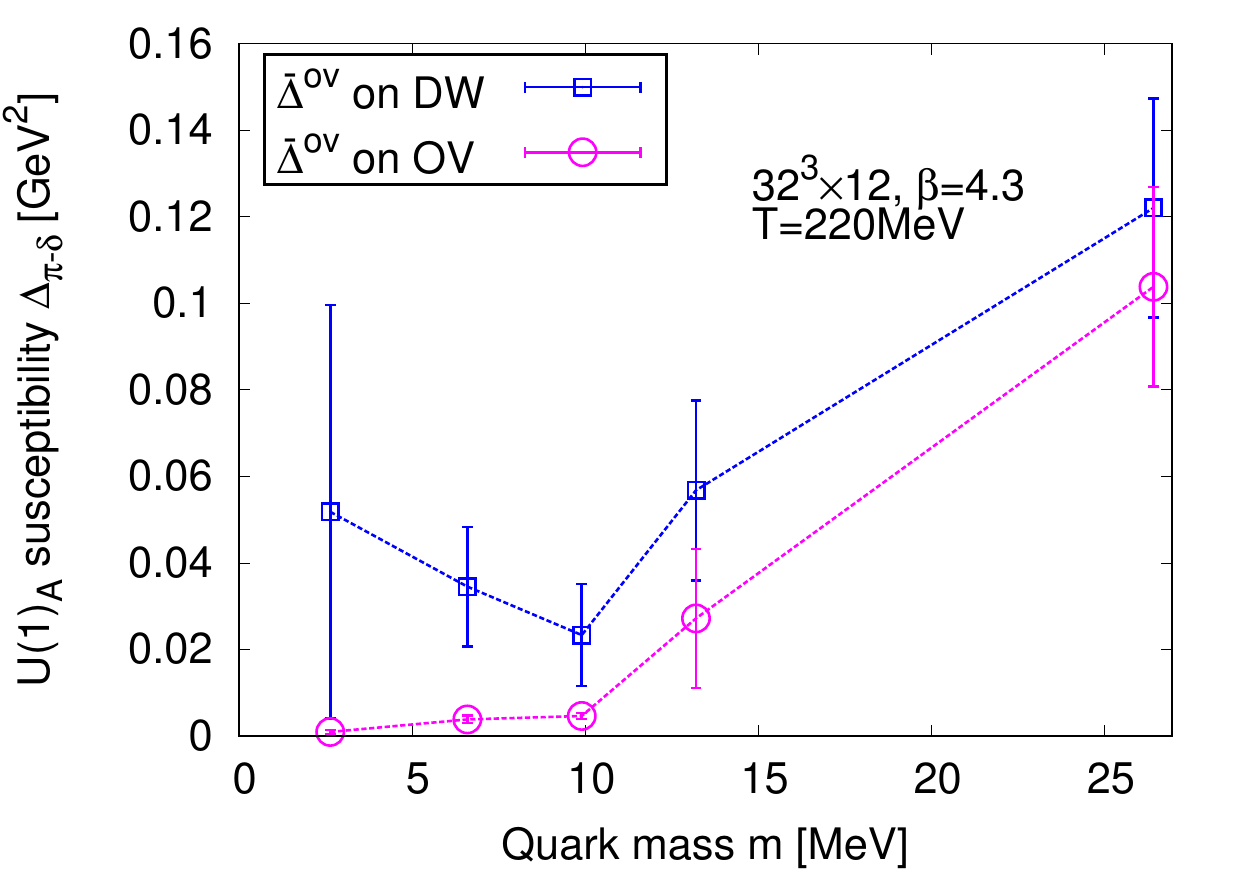}
        \end{center}
    \end{minipage}
    \begin{minipage}[t]{0.5\columnwidth}
        \begin{center}
            \includegraphics[clip, width=1.0\columnwidth]{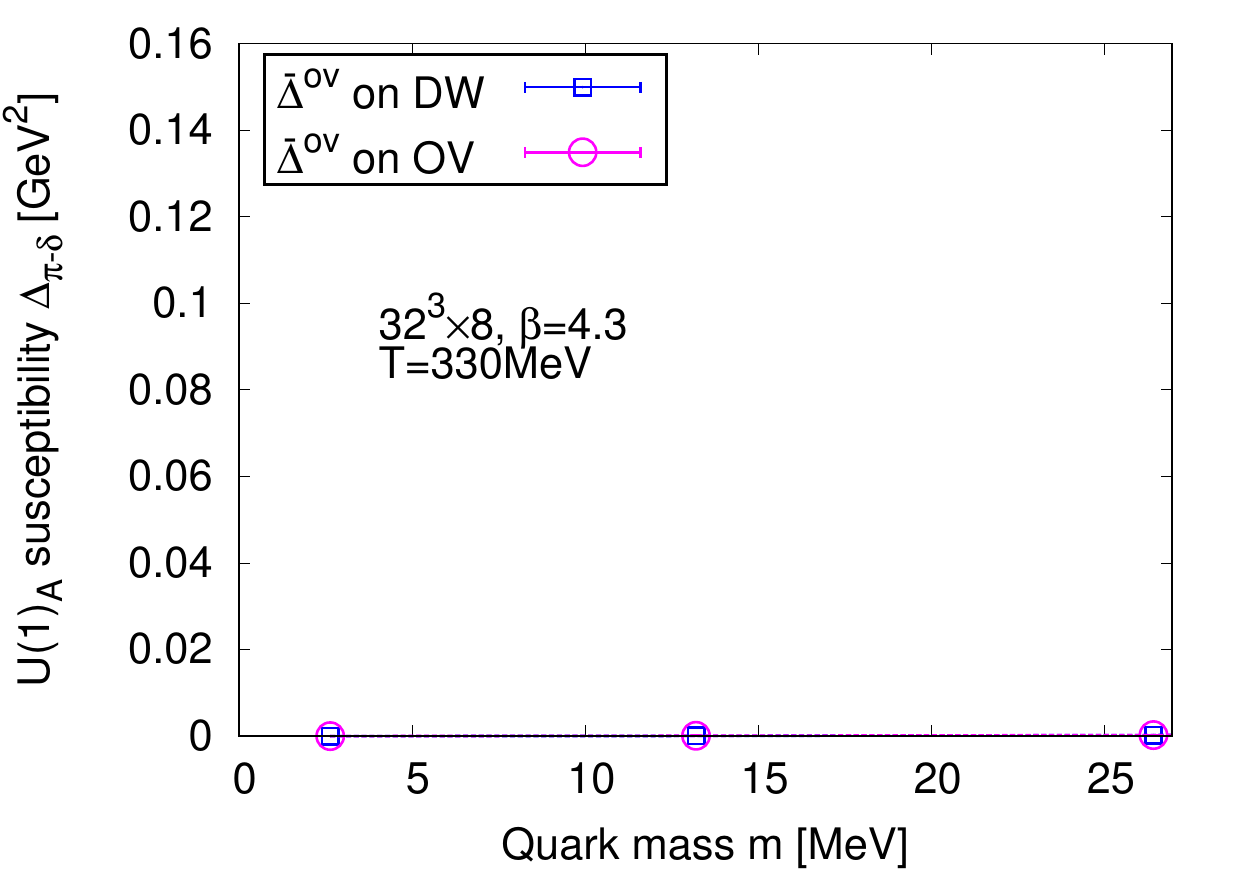}
        \end{center}
    \end{minipage}
    \caption{Quark mass dependences of $U(1)_A$ susceptibilities, $\bar{\Delta}_{\pi-\delta}^\mathrm{ov}$, from the eigenvalue density of the overlap Dirac operators on the M\"obius domain-wall (blue squares) and reweighted overlap (magenta circles) ensembles at $T=220 \, \mathrm{MeV}$ (left) and $T=330 \, \mathrm{MeV}$ (right).}
    \label{fig-2}
\end{figure}

\begin{figure}[tb!]
    \begin{minipage}[t]{1.0\columnwidth}
        \begin{center}
            \includegraphics[clip, width=0.5\columnwidth]{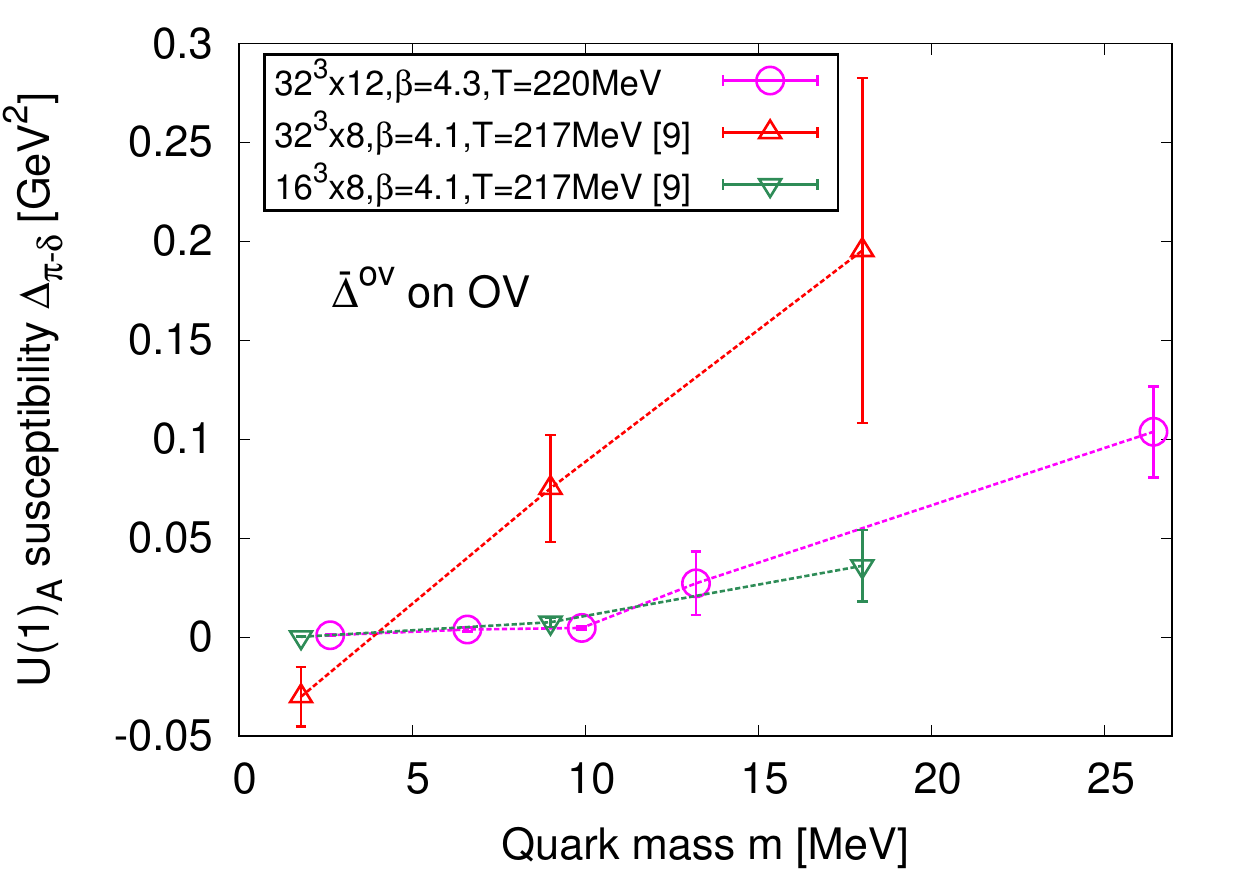}
        \end{center}
    \end{minipage}
    \caption{Quark mass dependences of $U(1)_A$ susceptibilities at $T \sim 220\, \mathrm{MeV}$ {\it by coarser lattices} with $a\sim 0.11 \, \mathrm{fm}$ (triangle points) in our previous work \cite{Tomiya:2016jwr}.
The results by the finer lattice (magenta circles) are the same as Fig.~\ref{fig-2}.}
    \label{fig-3}
\end{figure}

\subsection{$U(1)_A$ susceptibility}\label{subsec-3-2}
In the left panel of Fig.~\ref{fig-2}, we show the quark mass dependence of $U(1)_A$ susceptibility $\bar{\Delta}_{\pi-\delta}^{\mathrm{ov}}$.
Magenta circles and blue squares correspond to the results on the OV and DW ensembles, respectively.
Since $\bar{\Delta}_{\pi-\delta}^{\mathrm{ov}}$ on the DW suffers from fictitious modes by the partially quenched approximation, we expect that $\bar{\Delta}_{\pi-\delta}^{\mathrm{ov}}$ on the OV is closer to the continuum limit.
As expected, at a small quark mass $m \sim 3 \, \mathrm{MeV}$, $\bar{\Delta}_{\pi-\delta}^{\mathrm{ov}}$ on the OV nearly vanishes.
Therefore, in the chiral limit, $m \to 0$, we expect that the $U(1)_A$ symmetry is restored.
Furthermore, in the region around $m\sim 10 \, \mathrm{MeV}$, we find a sudden increase of $\bar{\Delta}_{\pi-\delta}^{\mathrm{ov}}$.
This behavior suggests the existence of a ``critical mass'' as suggested in Ref.~\cite{Aoki:2012yj}.
For the large quark mass region, $\bar{\Delta}_{\pi-\delta}^{\mathrm{ov}}$ has a large value, implying that the $U(1)_A$ symmetry is clearly broken.
Note that the results on the DW ensembles are consistent with those on the OV, for large $m$, and the partially quenched approximation works relatively well.

In Fig.~\ref{fig-3}, we show the results at the almost same temperature, $T=217 \, \mathrm{MeV}$, shown by triangle points, but on a coarser lattice ($a\sim 0.11 \, \mathrm{fm}$) in Ref.~\cite{Tomiya:2016jwr}.
Although the coarser lattice leads to larger violation of the GW relation for M\"obius DW fermion \cite{Cossu:2015kfa}, we found that $\bar{\Delta}_{\pi-\delta}^{\mathrm{ov}}$ is suppressed for the small quark mass region.
Therefore, both the finer ($a\sim 0.075 \, \mathrm{fm}$) and coarser lattices lead to the similar suppression of  $\bar{\Delta}_{\pi-\delta}^{\mathrm{ov}}$.

In the right panel of Fig.~\ref{fig-2}, we show the results at $T=330 \, \mathrm{MeV}$ which is much larger than $T_c$.
At this temperature the $U(1)_A$ susceptibility is highly suppressed at all the quark mass parameters we studied.
For instance, at $m \sim 27 \, \mathrm{MeV}$, $\bar{\Delta}_{\pi-\delta}^{\mathrm{ov}}$ after the reweighting is $\sim 10^{-4} \, \mathrm{GeV^2}$.
This behavior comes from the suppression of Dirac low modes by the appearance of a gap in the Dirac spectra at higher temperature.

\section{Conclusion and outlook}\label{sec-4}
In this study, we investigated the $U(1)_A$ symmetry above the critical temperature from numerical simulations in two-flavor lattice QCD.
The quark mass dependence of the $U(1)_A$ susceptibility at $T=220 \, \mathrm{MeV}$ implies the restoration of $U(1)_A$ symmetry in the chiral limit, which is consistent with the predictions of \cite{Aoki:2012yj,Kanazawa:2015xna}.
The zero modes of the overlap Dirac spectra are closely related to the topological susceptibility, which is reported in another talk \cite{Aoki:2017Lat}.
As other observables to see the restoration of $U(1)_A$, the spatial isovector meson correlators from the same gauge configuration are investigated in Ref.~\cite{Rohrhofer:2017grg}.

In the future, we will investigate $U(1)_A$ and topological susceptibilities at other temperature.
It is important to investigate the volume dependence and continuum limits in order to justify the restoration/violation of $U(1)_A$ symmetry.
The study of $N_f=2+1$ is a next step to be explored.
Previous studies \cite{Bazavov:2012qja,Buchoff:2013nra,Bhattacharya:2014ara,Dick:2015twa} suggested the violation of $U(1)_A$ symmetry, and we are interested in the results of the rigorous setup of this work. 

\section*{Acknowledgment}\label{sec-5}
Numerical simulations are performed on IBM System Blue Gene Solution at KEK under a support of its Large Scale Simulation Program (No. 16/17-14).
This work is supported in part by the Japanese Grant-in-Aid for Scientific Research (No. JP26247043), and by MEXT as ``Priority Issue on Post-K computer" (Elucidation of the Fundamental Laws and Evolution of the Universe) and by Joint Institute for Computational Fundamental Science (JICFuS).

\bibliography{lattice2017}

\begin{thebibliography}{20}

\bibitem{Cohen:1996ng}
T.D. Cohen, Phys. Rev. \textbf{D54}, R1867 (1996), \texttt{hep-ph/9601216}

\bibitem{Aoki:2012yj}
S.~Aoki, H.~Fukaya, Y.~Taniguchi, Phys. Rev. \textbf{D86}, 114512 (2012),
  \texttt{1209.2061}

\bibitem{Kanazawa:2015xna}
T.~Kanazawa, N.~Yamamoto, JHEP \textbf{01}, 141 (2016), \texttt{1508.02416}

\bibitem{Gross:1980br}
D.J. Gross, R.D. Pisarski, L.G. Yaffe, Rev. Mod. Phys. \textbf{53}, 43 (1981)

\bibitem{Pisarski:1983ms}
R.D. Pisarski, F.~Wilczek, Phys. Rev. \textbf{D29}, 338 (1984)

\bibitem{Cossu:2013uua}
G.~Cossu, S.~Aoki, H.~Fukaya, S.~Hashimoto, T.~Kaneko, H.~Matsufuru, J.I.
  Noaki, Phys. Rev. \textbf{D87}, 114514 (2013), [Erratum: Phys.
  Rev.D88,no.1,019901(2013)], \texttt{1304.6145}

\bibitem{Chiu:2013wwa}
T.W. Chiu, W.P. Chen, Y.C. Chen, H.Y. Chou, T.H. Hsieh (TWQCD), PoS
  \textbf{LATTICE2013}, 165 (2014), \texttt{1311.6220}

\bibitem{Brandt:2016daq}
B.B. Brandt, A.~Francis, H.B. Meyer, O.~Philipsen, D.~Robaina, H.~Wittig, JHEP
  \textbf{12}, 158 (2016), \texttt{1608.06882}

\bibitem{Tomiya:2016jwr}
A.~Tomiya, G.~Cossu, S.~Aoki, H.~Fukaya, S.~Hashimoto, T.~Kaneko, J.~Noaki,
  Phys. Rev. \textbf{D96}, 034509 (2017), \texttt{1612.01908}

\bibitem{Bazavov:2012qja}
A.~Bazavov et~al. (HotQCD), Phys. Rev. \textbf{D86}, 094503 (2012),
  \texttt{1205.3535}

\bibitem{Buchoff:2013nra}
M.I. Buchoff et~al., Phys. Rev. \textbf{D89}, 054514 (2014), \texttt{1309.4149}

\bibitem{Bhattacharya:2014ara}
T.~Bhattacharya et~al., Phys. Rev. Lett. \textbf{113}, 082001 (2014),
  \texttt{1402.5175}

\bibitem{Dick:2015twa}
V.~Dick, F.~Karsch, E.~Laermann, S.~Mukherjee, S.~Sharma, Phys. Rev.
  \textbf{D91}, 094504 (2015), \texttt{1502.06190}

\bibitem{Brower:2005qw}
R.C. Brower, H.~Neff, K.~Orginos, Nucl. Phys. Proc. Suppl. \textbf{153}, 191
  (2006), \texttt{hep-lat/0511031}

\bibitem{Brower:2012vk}
R.C. Brower, H.~Neff, K.~Orginos, Comput. Phys. Commun. \textbf{220}, 1 (2017),
  \texttt{1206.5214}

\bibitem{Ginsparg:1981bj}
P.H. Ginsparg, K.G. Wilson, Phys. Rev. \textbf{D25}, 2649 (1982)

\bibitem{Cossu:2015kfa}
G.~Cossu, H.~Fukaya, S.~Hashimoto, A.~Tomiya (JLQCD), Phys. Rev. \textbf{D93},
  034507 (2016), \texttt{1510.07395}

\bibitem{Fukaya:2013vka}
H.~Fukaya, S.~Aoki, G.~Cossu, S.~Hashimoto, T.~Kaneko, J.~Noaki (JLQCD), PoS
  \textbf{LATTICE2013}, 127 (2014), \texttt{1311.4646}

\bibitem{Aoki:2017Lat}
Y.~Aoki, S.~Aoki, G.~Cossu, H.~Fukaya, S.~Hashimoto, K.~Suzuki,
  \emph{{Topological Susceptibility in $N_f = 2$ QCD at Finite Temperature}},
  in \emph{Proceedings, \href{http://inspirehep.net/record/1425631}{35th
  International Symposium on Lattice Field Theory (Lattice2017)}: Granada,
  Spain}, to appear in EPJ Web Conf.

\bibitem{Rohrhofer:2017grg}
C.~Rohrhofer, Y.~Aoki, G.~Cossu, H.~Fukaya, L.{\relax Ya}. Glozman,
  S.~Hashimoto, C.B. Lang, S.~Prelovsek, Phys. Rev. \textbf{D96}, 094501
  (2017), \texttt{1707.01881}

\end{thebibliography}

\end{document}